# Inverted C-Shaped Slots Loaded Exponential Tapered Triple Band Notched Ultra Wideband (UWB) Antenna

Olaoluwa A. Adegboye
Department of Electrical, Electronic and Computer Engineering,
University of Uyo, Nigeria,

Kufre M. Udofia,
Department of Electrical, Electronic and Computer Engineering,
University of Uyo, Nigeria.

Akaninyene Obot,
Department of Electrical, Electronic and Computer Engineering, University of Uyo, Nigeria.

*Abstract*— This research presents a simple strategy for designing an exponentially tapered, triple-notched ultrawideband antenna. The antenna's microstrip line feed and radiating patch are matched using an exponential tapered transformer. This method inserts antenna notch elements, by cutting two inverted C-shaped slots in the radiating patch; frequency rejection can be achieved for WI-MAX and wireless LAN. The X-band is rejected by etching a U-shaped slot in the feedline. When embedding the notch elements, cross-coupling was minimized. The desired antenna was designed, simulated, and measured. The measured results and graphs show that our proposed design is reliable. This band-notched antenna rejects 3.5 GHz (Wi-MAX band, 3.3 to 3.7 GHz), 5.5 GHz (WLAN 2 band, 5.15 to 5.825 GHz), and 7.5 GHz (for satellite downlink X - band-7.25 GHz to 7.75 GHz). The proposed antenna meets UWB design requirements.

## I. INTRODUCTION

Existing licensed narrowband wireless interoperability microwave access WiMAX (3.3-3.7 GHz), wireless local area network -WLAN (5.15-5.85 GHz), and X-band satellite downlink (7.25-7.75 GHz), etc., can cause interference with the FCC's licensed ultrawideband (UWB) spectrum (Allen et al., 2006) of 3.1-10.6 GHz. Most of these narrow-band interference mitigation strategies use slots (Dong, Li, & Deng, 2017), parasitic resonators, and complementary split ring resonators (CSRR) (El Omari El Bakali, Zakriti, Farkhsi, Dkiouak, & El Ouahabi, 2021) on the radiator or feed section of the antenna to accomplish single/dual/triple notches. For the most part, printed monopole antennas are used as the reference antenna in frequency-notched UWB antennas because of their omnidirectionality and ease of design and manufacture. Tapered slot antennas (TSAs) have been suggested for usage in UWB applications due to its broad operational bandwidth, planar construction, extremely flat gain profile, and directional nature. TSAs can be excited in a variety of ways, including by using a linearly tapered profile, an exponentially tapered profile, or an antipodal arrangement. Despite this, UWB antennas need reliable notch functions regardless of the number of consecutive slots etched on the radiating patch.

The study of a CPW feed UWB antenna with a trapezoidal-shaped tuning stub, a small aperture area of 12 mm x 23 mm2, and periodic open-end stubs (POES) inserted into the inside edge of the rectangular aperture (Dastranj & Bahmanzadeh, 2020).

WLAN interference signals between 5 and 6 GHz were rejected using one of three notch filter designs, including a slot integrated on the patch, ground, and conductor-backed plane structure. UWB antenna with band-notch characteristic and CPW feeding was shown (Rani et. al, 2015). To increase the operational impedance bandwidth (2.38-10.6) GHz and the notch bandwidth, two inverted L-shaped slots were cut into the CPW ground. The impedance bandwidth was 2.38 GHz to 10.6 GHz at a VSWR 2, but the band between 5 and 6 GHz was rejected. The notch band feature was created by inserting a tuning stub in the middle of the fork-like patch.

Subbarao and Raghavan's (2013) suggested UWB antenna has a return loss of less than -10 dB throughout a frequency range of 3.1 GHz to 11 GHz. To prevent crosstalk between WLANs and HIPERLAN/2, a pot-shaped radiating patch with a C-shaped slit cut into its bottom was printed (5.1- 5.85 GHz). Except for the notched band, the operational bandwidth has a group delay of less than 1 ns from the antenna.

It was discovered that a small UWB printed slot antenna can support not only the standard Bluetooth frequency range but also an additional frequency band (5 GHz for WiMAX and 5.8 GHz for WLAN). Gurdeep and Urvinder (2021) suggested a stepped stub structure using the CPW feed approach to better fit the impedance of the circuit. The antenna had actual measurements of 32 mm x 29 mm. The antenna supported the 802.11b WLAN band in addition to the UWB bandwidth of 1 GHz-11.6 GHz. A stair-shaped slot put into the radiating patch's focal point was used to illustrate the band-notch function from 5.1 GHz to 5.9 GHz.

In this research, we present a novel design for an ultra-wideband (UWB) antenna that exhibits frequency notch characteristics and has an exponentially tapering profile. The primary antenna is a miniature band-notched circular patch antenna excited by a microstrip line. With the aid of an exponentially tapered section, the microstrip line is matched with the patch to enable distributed impedance over the UWB band. Besides the U-shaped slot etched at the antenna's narrowing tip, the patch is filled with a pair of inverted C-shaped slots. In (Saeidi, Ismail, Wen, Alhawari, & Mohammadi, 2019), an Archimedean spiral-shaped defect is embedded on the circular open-circuit stub of the microstrip to slot-line transition to create a frequency notch in the UWB band of a TSA. By loading two microstrip resonators and two varactor diodes, the authors of (Abbas, Hussain, Lee, Park, &

Kim, 2021) show that the notch frequency of a shortened version of the Vivaldi may be dynamically adjusted. As of 2020, SRR have been employed in both printed UWB monopole antennas and pyramidal horns ("Horn Antennas Loaded with Metamaterial for Satellite Band Application") (Jairath, Singh, Jagota, & Shabaz, 2021).

The proposed antenna features notches at 3, 5, and 7 GHz to accommodate WiMAX, WLAN, and X band, respectively. A circular patch antenna with ultra-wideband (UWB) capabilities is used in the design. FR-4, with a permittivity of 4.4, is the material of choice for this application.

## II. PROPOSED ANTENNA MODEL

The base antenna and the triple frequency notched antenna design are shown in Fig.1: 1(a) . The suggested antenna is fabricated on FR-4 substrate with relative permittivity, $\varepsilon_r=4.4$ and thickness $h=1.6mm$ in order to get better notching results than the preceding structure. In comparison with the preceding antenna structure, here the C-shaped Slots split gap($n_d$) is the same for the three slots. The prospective design is shown in Fig.1 (b) and design parameters are described in Table I. A circular patch of radius R is fed by a microstrip line of length $L_f$, width $W_f$, and impedance $Z_O$. The microstrip line is matched with the patch of impedance $Z_L$ by employing an exponential transformer whose exponential factor is az. The ground plane ($L_g$) of the proposed antenna is shortened and a square cut is applied to utilize the coupling between radiator and ground plane and to further increase the bandwidth. In the exponential taper line, the natural logarithm of taper line's characteristic impedance varies linearly from $Z_L$ to $Z_0$. The exponential taper has the form given in equation (1) and it is related to the reflection coefficient by the form in equation (3):

$$Z_i(z) = Z_o e^{az} \quad (1)$$
$$a = \frac{1}{L}\ln\frac{Z_L}{Z_o} \quad (2)$$

Reflection coefficients are given by equation (3):

$$\Gamma = \frac{\ln\frac{Z_L}{Z_o}}{2} e^{-i\beta l}\frac{\sin\beta l}{\beta l} \quad (3)$$

The bandwidth of a tapered line will typically increase as the length L increases.

TABLE 1: DESIGN PARAMETERS FOR THE BASE ANTENNA

| Design Parameter | Values | Optimized Values |
|---|---|---|
| **Substrate Dimensions** | | |
| Length of Substrate ($L_s$) | 31.969mm 31.969mm | 32mm |
| Width of Substrate ($W_s$) | 22.968mm 22.968mm | 22mm |
| Substrate height (h) | 1.6mm 1.6mm | 1.6mm |
| **Patch Dimensions** | | |
| Effective radius ($R_p$) | 6.553 mm | 5.8mm |
| Dielectric Constant ($\varepsilon_r$) | 4.4 | 4.4 |
| Patch thickness(t) | 0.0036mm | 0.0036mm |
| **Feedline Dimension** | | |
| Feed length | 5.42mm | |
| Feed Width | 3.13mm | 2mm |
| Exponential factor (az) | 2.37 | 2.37 |
| **Ground Plane** | | |
| Length of Ground ($L_g$) | 5.3mm | 5.3mm |
| Width of Ground ($W_g$) | 23mm | 23mm |

The band-notched functions of the antenna are achieved by etching two inverted C-shaped slots on the patch and a single U-shaped slot on the exponential matching section. The slots' length is one half of the guided wavelength at the respective notched and frequency (Gouda and Yousef, 2012). The length of each slot can be calculated using Equation (4)

$$f_{slot} = \frac{c}{2L_{slot}\sqrt{\varepsilon_{eff}}} \quad (4)$$

Where $c$ is the speed of light, $L_{slot}$ is the length of the slot, and $\varepsilon_{eff}$ is the effective dielectric constant of the substrate which is evaluated by equation (6) substrate. Taking slot frequency to be 3.5 GHz ($f_{slot1}$), 5.5 GHz ($f_{slot2}$) and 7.5 GHz ($f_{slot3}$), respectively, the equivalent slot length is evaluated below by rearranging Equation (4) to form Equation (5) and given in Table 2

$$L_{slot} = \frac{c}{2f_{slot}\sqrt{\varepsilon_{eff}}} \quad (5)$$

$$\varepsilon_{eff} = \frac{\varepsilon_r+1}{2} \quad (6)$$

TABLE 2: DESIGN PARAMETERS FOR NOTCH BANDS

| Slot No. ($F_r$ (GHz)) | Size (mm) | Opt. Size (mm) | Width (mm) | Length (mm) |
|---|---|---|---|---|
| Slot 1 (3.5) | 26.34 | 28 | 8 | 7 |
| Slot 2 (5.5) | 19.3 | 19 | 6 | 4.5 |
| Slot 3 (7.5) | 12.29 | 13.2 | 1.6 | 5.8 |

## III. SIMULATED RESULTS AND ANALYSIS

The base antenna design is simulated, and Figure 2 shows the S-parameters. The antenna is seen to radiate below -10dB from 2.78 GHz to 11.97 GHz, covering the UWB frequency range of 3.2 – 10.6 GHz.

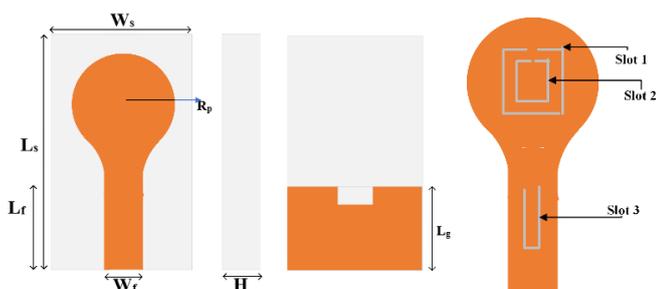

Fig.1. (a) Schematic of the base antenna (printed circular patch on FR-4) (b) Back view of the printed circular monopole antenna with shortened ground plane (c) Schematic of the proposed design with slots

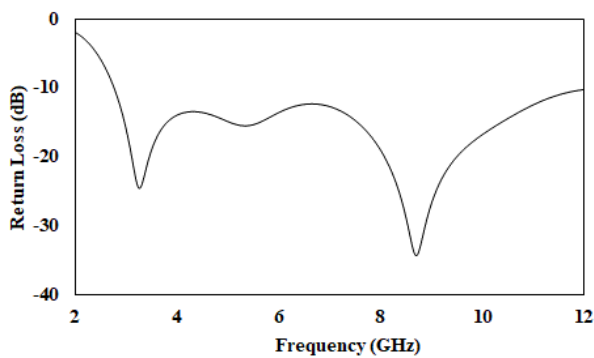

Fig. 2: Return Loss (S11) of the Circular Patch Antenna

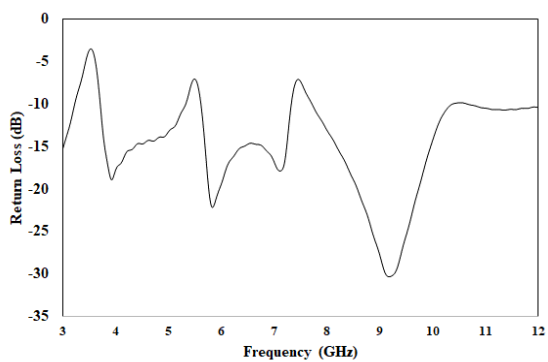

Fig. 3: Return loss curve of proposed antenna with Tri - band notch function.

The slots were also etched on the base antenna. On a FR-4 substrate with thickness h=1.6mm and dielectric constant $\varepsilon_r$ =4.4, a microstrip line fed circular patch UWB antenna with exponentially tapered match section for stable UWB impedance matching and inverted C-Slots for triple frequency notches is simulated. The simulation yields notches at the desired frequencies of 3.5GHz, 5.5GHz, and 7.5GHz. The simulated return loss characteristics of the antenna are shown in Fig. 3, and it can be seen that the signal at these three frequencies was reflected back. Figure 4 shows the simulated VSWR of an antenna with triple band-notched characteristics. At frequencies centered at 3.5 GHz (3.29 GHz to 3.75 GHz), 5.5 GHz (5.3 GHz - 5.7 GHz), and 7.5 GHz, the antenna has a voltage standing wave ratio greater than two (VSWR 2). (7.25 GHz – 7.75 GHz). Figure 5 depicts the gain of the proposed antenna. The maximum gain of the antenna is 3.2 dBi at 9 GHz, while the gain drops to about -3.3 dBi at notch band center frequency of 3.5 GHz, -1dBi at 5.5 GHz, and -1.1dBi at 7.5 GHz.

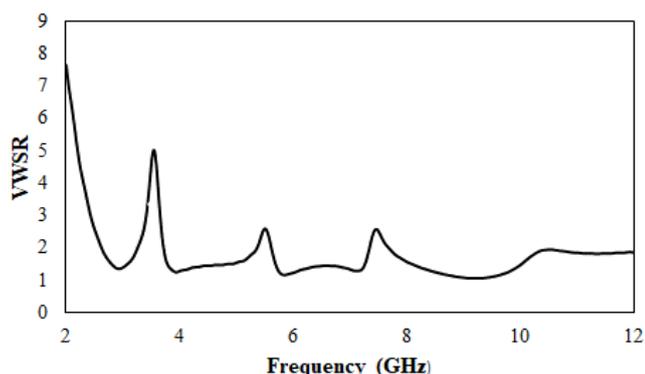

Fig. 4: Simulated VSWR curve of proposed antenna with Tri - band notch function.

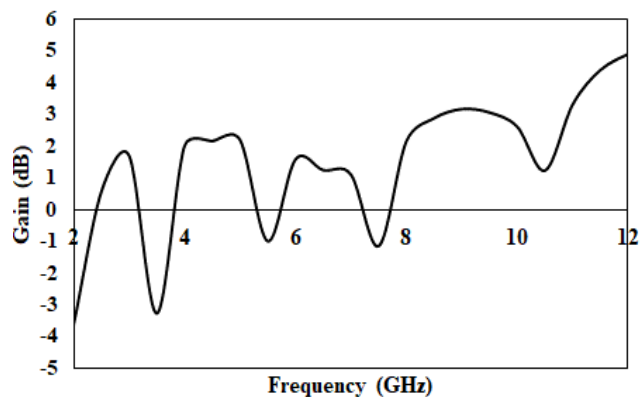

Fig. 5: Graph of gain against frequency for proposed antenna

IV. MEASURED RESULT

The final design of the exponentially tapered Triple Band – Notched UWB antenna as shown in Figure 6 was fabricated and measurements carried out in the Microwave Laboratory of the International Islamic University of Malaysia. The return loss and VWSR of the fabricated antenna is reported in this section. The graph of measured return loss against frequency of the fabricated antenna is presented in Figure 7. The measured impedance bandwidth of the proposed antenna is 2.7 - 11.3 GHz with three stop bands centered at 3.5, 5.5 and 7.5 GHz. As a result, the antenna is able to avoid interference with narrowband services of WiMAX, WLAN and also X band satellite communication. The measured results are in good agreement with the simulated ones with very little variation.

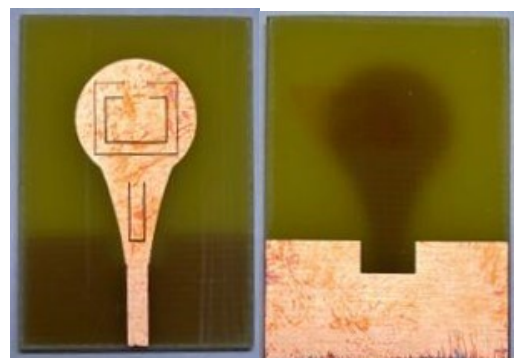

Fig. 6: Fabricated Triple Band Notched UWB Antenna
Source: IIUM Fabrication Laboratory

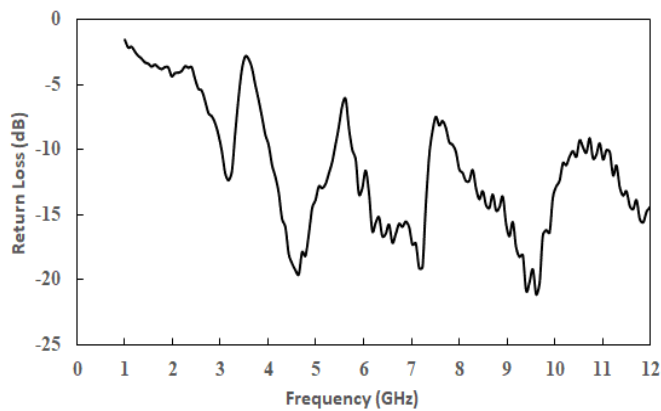

Fig. 7: Measured return loss (S11) against Frequency

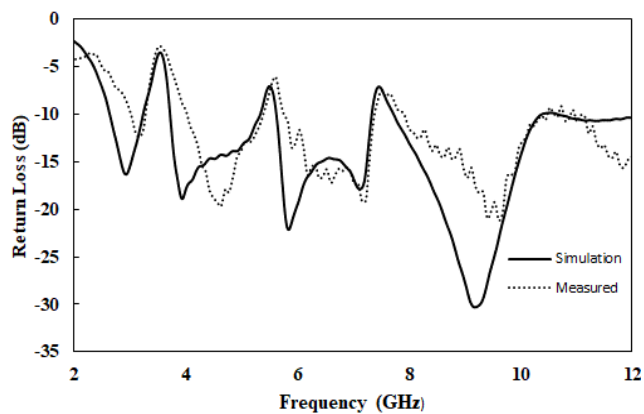

Fig. 10: Simulated and Measured Return Loss of Proposed Antenna

## V. COMPARISON OF SIMULATED AND MEASURED RESULTS

Figure 10 shows simulated and measured the proposed UWB antenna return loss. The proposed antenna has a 2.7 to 10.3 GHz impedance bandwidth at -10 dB S11, with stop bands centered at 3.5, 5.5, and 7.8 GHz. Since WiMAX, WLAN, and X band satellite use narrowband services, the antenna won't interfere. Measured and simulated results shows nearly the same performance. Fabrication flaws, improper SMA connector soldering, or the testing environment contribute to the observed disparity between simulated and measured responses.

In this paper, a circular patch antenna with triple band notches is designed. The antenna had stable FCC-defined UWB bandwidth. Return loss, VSWR, and realized gain, at notched band center frequency, were used to investigate the proposed design. At the notched band's center frequency, current concentration is high near the slot. We measured the performance of the three-slot antenna. Simulation and measurements agree on the antenna's radiation properties and gain suppression at notched bands. Each slot's center frequency can be adjusted independently of the others.

Table 4 compares the implemented antenna structure to UWB antenna design requirements.

TABLE 4: SUMMARY OF OUR DESIGN RESULTS COMPARED WITH UWB REQUIREMENTS

| Parameter | Requirements | Our Design results |
|---|---|---|
| VSWR Bandwidth | 3.1 - 10.6 GHz | 2.7 - 11.3 GHz |
| Radiation Efficiency | High (>70%) | 86% |
| Phase | Linear; Constant Group Delay | Constant |
| Radiation Pattern | Omni directional | Omni directional |
| Directivity and Gain | Low | 3.8dBi |
| Half Power Beamwidth Wide | (>60%) | 70 |
| Physial Profile | Small, Compact, Planar | Small, Compact, Planar |

## VI. CONCLUSION

This study designs and implements a triple-notch UWB antenna. First, a microstrip antenna is designed. The antenna is fed by microstrip line and matched to the patch with an exponential transformer to increase bandwidth. CST Microwave studio simulated the antenna. Slots on the radiating patch change the simple circular microstrip antenna. At specified frequencies, the goal is to disrupt the radiating patch's current distribution. Antenna parameters are tested. The antenna spans 2.75 to 11 GHz.

A circular UWB antenna with triple band-notches is demonstrated. Two C-shaped slots were etched on the radiating patch to achieve band- -notched characteristics for WiMAX (3.5 GHz) and WLAN (5.5 GHz). A U-shaped slot was etched on the feedline to obtain band-notched characteristics at Satellite Downlink X-Band (7.5 GHz) frequency. The antenna spans 2.8-11.2 GHz.

The presented antenna provides nearly omnidirectional patterns at 3-11 GHz. Due to fabrication flaws, SMA connectors, and dielectric constant non-uniformity, measured and simulated results varied.